\documentclass[preprint,prd,showpacs,floatfix]{revtex4}
 
\usepackage{amsmath,amsfonts,amssymb}
\usepackage[dvips]{graphicx}

\begin{document}

\title{Can the initial singularity be detected by cosmological tests?}
\author{Marek Szyd{\l}owski, \email{uoszydlo@cyf-kr.edu.pl}
W{\l}odzimierz God{\l}owski}
\affiliation{Astronomical Observatory, Jagiellonian University,
Krak{\'o}w, Poland}
\author{Adam Krawiec}
\affiliation{Institute of Public Affairs, Jagiellonian University,
Krak{\'o}w, Poland}
\author{Jacek Golbiak}
\affiliation{Department Philosophy of Nature and Philosophy of
Natural Sciences, Catholic University of Lublin, Lublin, Poland}

\begin{abstract}
In the presented paper we raised the question whether initial cosmological
singularity can be proved by cosmological tests. The classical
general relativity theory predicts the existence of singularity in the past
if only some energy conditions are satisfied. On the other hand the
latest quantum gravity applications to cosmology suggest the possibility of
avoiding the singularity and replacing it with a bounce. Bounce is the
moment in the evolution of the Universe when the Universe's size has minimum.
Therefore the existence of observationally detected bounce in past of
Universe could indicate the validity of the loop quantum gravity hypothesis
and nonexistence of initial singularity which is present in the classical
$\Lambda$CDM. We investigated the bouncing model described by the generalized
Friedmann-Robertson-Walker (FRW) equation in the context of the observations
of the currently accelerating universe. The distant type Ia supernovae data
are used to constraint on bouncing evolutional scenario where square of
the Hubble function $H^2$ is given by formulae
$H^2=H^2_0\left[\Omega_{m,0}(1+z)^{m}-\Omega_{n,0}(1+z)^{n}\right]$, where
$\Omega_{m,0}, \Omega_{n,0}>0$ are density parameters and $n>m>0$.
In this paper are
showed that the on the base of the SNIa data standard bouncing models can
be ruled out on the $4\sigma$ confidence level. After adding the cosmological
constant to the standard bouncing model (the extended bouncing model)
we obtained as the best-fit that the parameter $\Omega_{n,0}$ is equal zero
which means that the SNIa data do not support the bouncing term in the model.
The bounce term is statistically insignificant on the present epoch.
We also demonstrated that BBN offers the possibility of obtaining stringent
constraints of the extra term $\Omega_{n,0}$. The other observational test
methods like CMB and the age of oldest objects in the Universe are also used.
We use as well the Akaike informative criterion to select a model which fits
data the best and we concluded that bouncing term should be ruled out by
Occam's razor, which makes the big bang scenario more favorable then the 
bouncing scenario.
\end{abstract}

\pacs{98.80.Bp, 98.80.Cq, 11.25.-w}
 
\maketitle

\section{Introduction}
 
We are living in an age of high precision cosmology which offers a
possibility of testing exotic physics, which is obvious for the early
Universe \cite{Lahav:2004iy}. In this context the most important are BBN
constraints because the present Universe opens only small windows on the
exotic physics. The main aim of this paper is to discuss whether the initial
singularity can be checked against the astronomical observations. The question
of singularity cannot be answered directly, therefore we use two prototype
models based on the classical and quantum gravity theory. The first is
the $\Lambda$CDM which is a concordance model describing the evolution of
the Universe from the initial singularity (the big bang) driven by
the cold dark matter and the cosmological constant (dark energy). The second
is a bouncing model which appears in the context of quantum cosmology and
characterized by the lack of initial singularity. During its evolution, the
expansion phase is proceeded by the contraction phase at the bounce where
the scale factor assumes the minimum nonzero value.
 
We use some tests to discriminate between these two alternative models.
One of the most important tests applies the SNIa data to fit the cosmological
models. Recent measurements of type Ia supernovae observations suggest that the
universe is presently accelerating \cite{Riess:1998,Perlmutter:1999}. A dark 
energy component has usually been proposed as a source of acceleration 
mechanism \cite{Peebles:2003}. 
Many theoretical propositions have been suggested about these components.
However, the different effects arising from quantum fluctuation, spinning
fluid, etc. can also mimic dynamically the role the dark energy
which drives acceleration through an additional term in the Friedmann equation
\cite{Zhu:2003,Sen:2003,Godlowski:2003hf,Godlowski:2004,Godlowski:2004b,
Dabrowski:2004,Godlowski:2004c,Puetzfeld:2004,Padmanabhan:2003b,Choudhury:2005}.
Some of them give rise to the bounce. In many cases they prevailed
in the very early epoch but are very small in the present epoch. Therefore
it is very difficult to detect the existence of this component in the
present and those of the relatively close past (after CMB) observations of SNIa.
 
In the present work we investigate observational constraints on the evolutionary
scenario of the standard bouncing cosmological models defined as a class of
models for which the Hubble function $H$ and the scale factor $a$
are related by the formula
\begin{equation}
\label{eq:1}
H^2=H^2_0(\Omega_{m,0}x^{-m}-\Omega_{n,0}x^{-n})
\end{equation}
where $n>m>0$ and $x=\frac{a}{a_0}$ where the index zero denotes the
quantities evaluated in the present epoch, the parameter $\Omega_{n,0}$ is
called the bouncing term; and the density parameters satisfy the constraint
relation
\[
\Omega_{m,0} - \Omega_{n,0} = 1.
\]
While focus mainly on the constraints coming from SN Ia data and WMAP
observations, the complementary constraints coming from BBN and the age 
\cite{Feng:2005} of the
oldest high-redshift objects are also considered. We use the maximum
likelihood method to estimate the model parameters $m$, $n$ and $\Omega_{n,0}$.
Similarly we analyze the models with the additional parameter---the
cosmological constant. It is called the generalized bouncing model.
 
The proposition of the bounce type evolution of the early universe seems to be
very attractive not only from the point of view of the quantum description of
the early Universe because the expansion of the universe is accelerated
automatically due to the presence of the bouncing term.
 
The standard bouncing scenario predicts the acceleration around the bounce
with a transition to the deceleration epoch. The cosmological constant
brings this deceleration epoch to the end and a new acceleration epoch begins.
 
Therefore, these models can be proposed as the models of our Universe,
because they include the epoch of acceleration. However we show that the
influence of the bouncing term is insignificant in the present epoch. Therefore,
the data from the present epoch, such as the SNIa data, have not power to
rsider the model with bouncing term statistically significant.
So the $\Lambda$CDM model with big-bang scenario is strongly
favored by data over the model with the bounce.
 
By the application of standard Akaike criterion of the model selection we can
choose the $\Lambda$CDM model over the generalized bouncing model. We conclude
that the data fail to support the existence of the bouncing term. The bouncing 
term in the present epoch is insignificant and it is not possible to detect 
its influence by the use the latest SNIa data.
 
This fact justifies certain scepticism about the existence of the SNIa window
on exotic physics in the current epoch. However we cannot rule out other
models by testing them against the current data. It is also possible to 
investigate the differences in the predictions of these models for some 
earlier epoch.
 
For example the BBN epoch is a well tested area of cosmology. From this
analysis we gather that the extra term $\Omega_{n,0}x^{-n}$ causing the bounce
should be constrained to be sufficiently small during nucleosynthesis.
 
The organization of the text is the following. In section 2 the evolutionary
scenario of bounce FRW cosmologies is investigated by the use of dynamic
system methods. We show that they are structurally unstable due to the
presence of centers in the phase portraits. In section 3 we discuss the
constraints from SNIa data on the standard bouncing models. In section 4
we extend the bouncing models by introducing the cosmological constant and
then we study how these models fit the  current supernovae and WMAP data.
In section 5 we formulate conclusions.

\section{The bouncing models: basic equations}
 
The idea of bounce in FRW cosmologies appeared in Tolman's monograph devoted 
to cosmology \cite{Tolman:1934}. This idea was strictly connected with 
oscillating models \cite{Robertson:1933,Einstein:1931,Tolman:1931}.
At present oscillating models play an important role in the brane cosmology 
\cite{Steinhardt:2002kw,Shtanov:2002ek}. The FRW universe undergoing a bounce 
instead of the big-bang is also an appealing idea in the context of quantum 
cosmology \cite{Coule:2004qf}. The attractiveness of bouncing models comes from 
the fact that they have no horizon problem and they explain quantum origin 
of structures in the Universe \cite{PintoNeto:2004wf,Barrow:2004ad,Salim:2005}. 
Molina-Paris and Visser and later Tippett \cite{Molina-Paris:1998tx,Tippett:2004xj} 
characterized the bouncing models by the minimal condition under which the 
present universe arises from a bounce from the previous collapse phase 
(the Tolman wormhole is different name for denoting such a type of evolution). 
The violation of strong energy condition (SEC) is in general a necessary (but 
not sufficient) condition for bounce to appear. For closed models it is 
sufficient condition and none of other energy condition need to be violated 
(like null energy condition (NEC): $\rho+p \ge 0$, week energy condition 
(WEC): $\rho \ge 0$ and $\rho + p \ge 0$, dominant energy condition (DEC): 
$\rho \ge 0$ and $\rho \pm p \ge 0$ energy conditions can be satisfied).
 
We can find necessary and sufficient conditions for an evolutional path with 
a bounce by analyzing dynamics on the phase plane ($a, \dot{a}$), where $a$ is 
the scale factor and dot denotes differentiation with respect to cosmological 
time. We understand the bounce as in \cite{Molina-Paris:1998tx,Tippett:2004xj}, 
namely there must be some moment say $t=t_{\text{bounce}}$ in evolution of 
the universe at which the size of the universe has a minimum, 
$\dot{a}_{\text{bounce}}=0$ and $\ddot{a} \ge 0$. This weak inequality 
$\ddot{a}\ge 0$ is enough for giving domains in the phase space occupied by 
trajectories with the bounce. Let us consider the dynamics of the FRW 
cosmological models filled by perfect fluid with energy density $\rho$ and 
pressure $p$ parameterized by the equation of state in the general form
\begin{equation}
p=w(a)\rho.
\label{eq:2}
\end{equation}

The basic dynamical system constitutes two equations
\begin{align}
\frac{\ddot{a}}{a} &= -\frac{1}{6}(\rho+3p)
\label{eq:3a} \\
\dot{\rho} &= -3H(\rho+p)
\label{eq:3b}
\end{align}
Equation (\ref{eq:3a}) is the Rauchadhuri equation while equation (\ref{eq:3b})
is the conservation condition. If the equation of state is postulated in the 
form (\ref{eq:2}) then from (\ref{eq:3b}) we obtain
\begin{equation}
\rho=\rho(a)=\rho_0 a^{-3} \exp{\left(-3 \int^a \frac{w(a)}{a}da\right)}
\label{eq:4}
\end{equation}
It is interesting that dynamics of the model under consideration can be
represented in the analogous form to the Newtonian equation of motion
\begin{equation}
\ddot{a}=-\frac{\partial V}{\partial a}
\label{eq:5}
\end{equation}
where $V= -\frac{\rho a^2}{6}$ plays the role of potential function for the
FRW system. Therefore different cosmological models are in a unique way
characterized by the potential function $V=V(a)$ and we can write down the
Hamiltonian for the fictitious particle-universe moving in the one-dimensional 
potential as
\begin{equation}
{\cal H} =\frac{p^2_a}{2}+V(a), \qquad p_a=\dot{a}
\label{eq:6}
\end{equation}
 
It is useful to represent eq. (\ref{eq:5}) in the form of dynamical system
\begin{equation}
\dot{x}=y, \qquad
\dot{y}=-\frac{\partial V}{\partial x},
\label{eq:7}
\end{equation}
where we denote $x=a$, $y=\dot{a}$ and system (\ref{eq:7}) has the first
integral in the form
\begin{equation}
\frac{y^2}{2}+V(x)=-\frac{k}{2}
\label{eq:8}
\end{equation}
where $k$ is the curvature index.
 
The critical points of the system (\ref{eq:7}) if exist are: $y_0=0$,
$(\frac{\partial V}{\partial x})_{x_0}=0$, i.e. they are always static
critical points located on $x$-axis. The form of first integral (\ref{eq:8})
defines the algebraic curves in the phase plane $(a,\dot{a})$ on which lies
solutions of the system. This solutions are in two types: regular is 
represented by trajectories or singular is represented by singular solutions 
for which the right-hand side of (\ref{eq:7}) are null (or 
$V(x_0)=-\frac{k}{2}$ for nonflat models). Note that the bouncing points are 
intersections points of trajectories situated in the region of the 
configuration space in which $\frac{\partial V}{\partial a} \le 0$, i.e. 
$V(a)$ is a decreasing function of $a$ or has extrema.
It is well known that the systems in the form (\ref{eq:7}) have only critical
points of two admissible types: centres if $(\frac{d^2 V}{d x^2})_{x_0}>0$
or saddles in opposite case if $(\frac{d^2 V}{d x^2})_{x_0}<0$. 
Therefore all trajectories with bounce intersect an $x$-axis and then they are
situated on the right side from the critical point at which $\dot{a}=0$ and
$\ddot{a} \ge 0$. The critical points are represented by points as well as
by separatrices of the saddle point. In others words bouncing trajectories are
represented by such trajectories in the phase plane which are passing through 
the $x$-axis in such a direction that they always belong to the accelerating 
region (in the neighborhood of bounce). Of course it is only possible if the 
SEC is violated.
 
Let us consider some prototype of bouncing models given by the Friedmann
first integral in the form
\begin{equation}
H^2=\frac{A}{a^m}-\frac{B}{a^n},
\label{eq:9}
\end{equation}
where $A$, $B$ are positive constants and $n>m$, $H=(\ln{a})^{.}$ is the
Hubble function and a dot denotes differentiation with respect to
cosmological time $t$.
 
It is convenient to rewrite (\ref{eq:9}) to the new form
\begin{equation}
H^2=H^2_0(\Omega_{m,0}x^{-m}-\Omega_{n,0}x^{-n}),
\label{eq:10}
\end{equation}
where $\Omega_{m,0}$, $\Omega_{n,0}$ are density parameters for
noninteracting fluids which give some contributions to right-hand sides
of eq.~(\ref{eq:9}). We define density parameters
$\Omega_{m,0}=\frac{3Aa^{-m}}{3H^2_0}$, $\Omega_{n,0}=\frac{3Ba^{-n}}{3H^2_0}$,
where an index ``$0$'' means that corresponding quantities are evaluated at 
the present epoch, $x=\frac{a}{a_0}$ is the scale factor expressed in the 
units of its present value $a_0$.
 
After differentiation the both sides of (\ref{eq:10}) with respect to the
reparameterized time variable $\tau \colon t \to \tau$, $|H_0|dt=d\tau$ we 
obtain
\begin{equation}
\frac{\ddot{x}}{x}=\frac{1}{2}\left(\Omega_{m,0}\left(2-m\right)x^{-m}+
\Omega_{n,0}\left(n-2\right)x^{-n}\right)
\label{eq:11}
\end{equation}
 
If we consider the generalization of the bouncing models with the cosmological
constant then in both equations (\ref{eq:10}) and (\ref{eq:11}) the parameter 
$\Omega_{\Lambda,0}$ should be added to their right-hand sides.
 
Note that the bouncing models can be treated as the standard FRW models 
with two noninteracting fluids with energy density and pressure in the form 
\begin{align*}
\rho = \rho_{m} + \rho_{n} = 3H_{0}^{2} \Omega_{m,0}x^{-m} -
3H_{0}^{2}\Omega_{n,0}x^{-n} \\
p = \left( -1 + \frac{m}{3} \right) \rho_{m}
+ \left(-1 + \frac{n}{3} \right) \rho_{n} \qquad
\rho_{m}>0, \rho_{n}<0.
\end{align*}
The curvature term as well as cosmological constant term can be obtained 
in an analogous way by putting $m=2$ or $m=0$, respectively.

If we postulate that the present universe is accelerating, i.e. $\ddot{x}>0$
at $x=1$ then in the general case with the cosmological constant we obtain
the following condition
\begin{equation}
\Omega_{m,0}\left(2-m\right)+
\Omega_{n,0}\left(n-2\right)+2\Omega_{\Lambda,0}>0
\label{eq:12}
\end{equation}
Because relation (\ref{eq:12}) is validate any time, the substitution
$H=H_0$ and $x=1$ to (\ref{eq:12}) gives constraint
\begin{equation}
\Omega_{m,0}-
\Omega_{n,0}+\Omega_{\Lambda,0}=1
\label{eq:13}
\end{equation}
Let us now concentrate on the standard bouncing models (SB) without the
cosmological term. Then from (\ref{eq:12}) including constraints
(\ref{eq:13}) we obtain the sufficient condition for acceleration at present
\begin{equation}
\Omega_{m,0}(n-m)>2-n,
\label{eq:14}
\end{equation}
where for the case of $n=2$, $k=1$ $\Omega_{m,0}>0$ is only required for
present acceleration.
 
If only $\Omega_{n,0}$ is larger then $\Omega_{m,0}$ the bouncing universe is 
presently accelerating for any $m$, $n$ parameters. It is worthy to mention 
that condition (\ref{eq:14}) is minimal qualitative information about 
acceleration and the rate of this acceleration is required for explanation 
SNIa data.
 
From the definition(\ref{eq:10}) one can obtain the domain admissible for
motion of the bouncing models
\begin{equation}
D=\{x\colon x \ge x_{\text{b}} \qquad \mathrm{where} \qquad
x_{\text{b}}=\left(\frac{\Omega_{n,0}}{\Omega_{m,0}}\right)^{\frac{1}{n-m}}\}.
\label{eq:15}
\end{equation}
From (\ref{eq:10}) the potential function $V(x)$ in the particle-like
description can be determined 
\begin{equation}
V(x)=-\frac{\rho_{eff}x^2}{6H^2_0}=-\frac{1}{2}\Omega_{eff}(x)x^2
\label{eq:16}
\end{equation}
where effective density parameter
$\Omega_{eff}=\Omega_{m,0}x^{-m}-\Omega_{n,0}x^{-n}$.
The acceleration region in the phase plane  can be determined in term of
potential function, namely if
\begin{equation}
\frac{dV}{da}<0
\label{eq:17}
\end{equation}
then universe is accelerating.
 
From (\ref{eq:16}) we obtain result that at the bounce moment
\begin{equation}
\ddot{x}=-\left(\frac{dV}{dx}\right)_{x_b}=\frac{1}{2} \Omega_{m,0}
\left(\frac{\Omega_{m,0}}{\Omega_{n,0}}\right)^{\frac{m}{m-n}}\left(n-m\right)
\label{eq:18}
\end{equation}
which indicate that bouncing models  defined by equation (\ref{eq:10})
at the bounce are in accelerating phase for any ranges of model parameter
$m,n,\Omega_{m,0},\Omega_{n,0}$. Because $a_b$ is larger then
$a_0 \colon (\frac{dV}{da})_{a_0}=0$ the  bouncing universe stay still in
the accelerating region.
 
From eq.~(\ref{eq:18}) we obtain that the universe start to accelerate at
the point $x=x_0$ such that
\begin{equation}
x_0=\left(\frac{\Omega_{m,0}\left(m-2 \right)}
{\Omega_{n,0}\left(n-2 \right)}\right)^{\frac{1}{m-n}}
\label{eq:19}
\end{equation}
where a positive value of $(m-2)(n-2)$ is required.
Note that in any case
\begin{equation}
x_0<x_b
\label{eq:20}
\end{equation}
i.e., the start of acceleration proceeds the bounce. The value of $x_0$ 
determine the location of the critical point of the dynamical system
$\dot{x}=y, \dot{y}=-\frac{\partial V}{\partial x}$ on $x$-axis. The sign of
the second derivative of potential function determines the type of critical
points (centre or saddle) because the eigenvalues of the linearization matrix
of the system satisfy the characteristic equation
\begin{equation}
\lambda ^2+\left(\frac{\partial ^2 V}{\partial x^2}\right)\left(x_0\right)=0.
\label{eq:21}
\end{equation}
From (\ref{eq:16}) we obtain
\begin{equation}
\left(\frac{\partial^2 V}{\partial x^2}\right)\left(x_0\right)=
-\frac{1}{2}\Omega_{n,0}x_0^{-n}
\left[(2-m)(1-m)-(2-n)(1-n)\right].
\label{eq:22}
\end{equation}
 
Therefore if $(m,n)$ belong to the interval $(3/2, \infty)$ then we have centres,
while if $(m,n)$ belong to interval $(-\infty,3/2)$ we obtain saddles.
 
\begin{figure}
\includegraphics[width=0.70\textwidth]{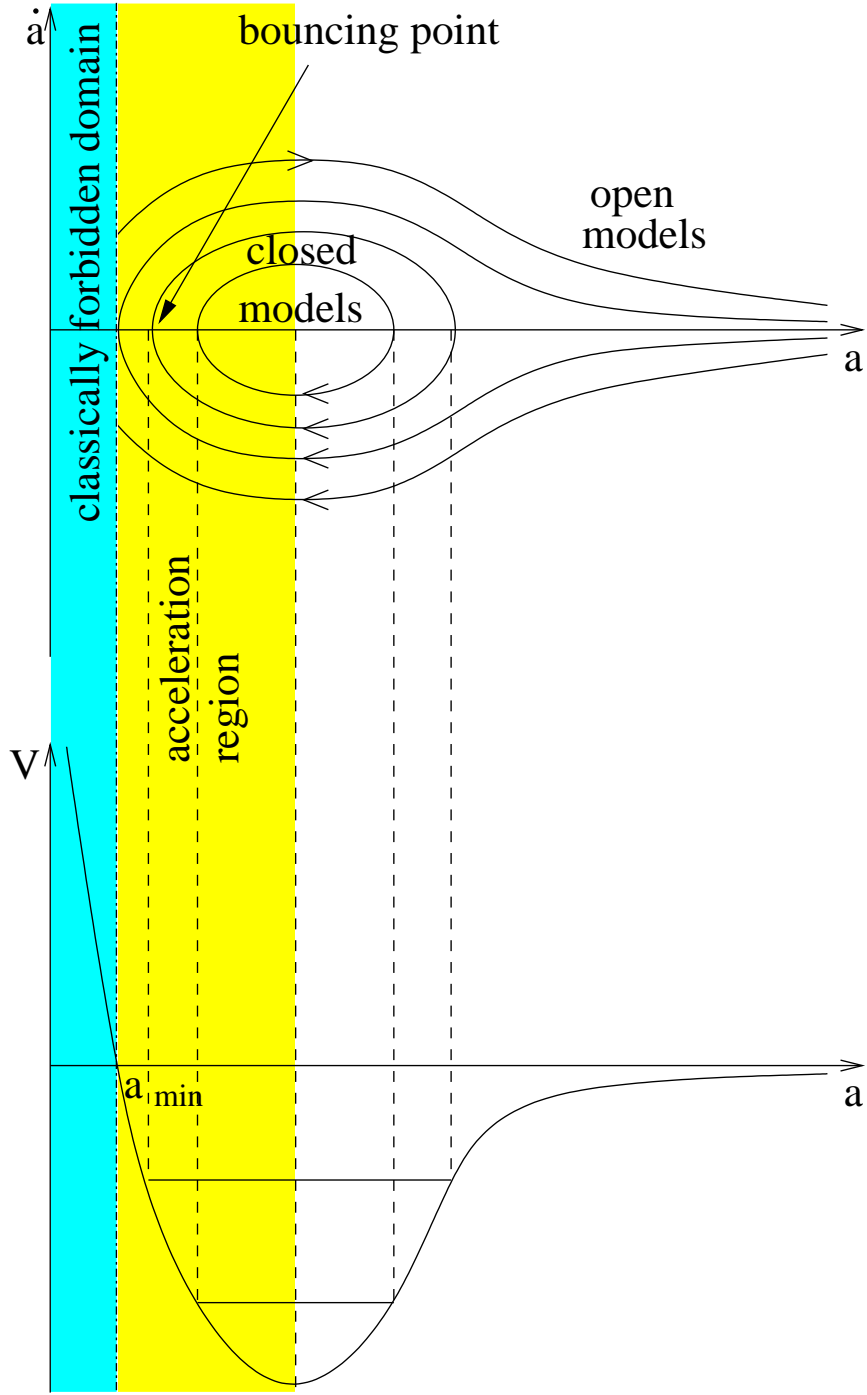}
\caption{The phase portrait and the diagram of the potential function for
BM model (all case from Table \ref{tab:1}). The minimum of the potential
function corresponds to a centre on the phase plane. The acceleration region
is located on the right from the $a_{\text{min}}$}
\label{fig:1}
\end{figure}

Let us concentrate, for example, let us concentrate on the case of $m=3$ 
(dust matter). Then
\begin{equation}
\left(\frac{\partial^2 V}{\partial x^2}\right)\left(x_0\right)=
\frac{1}{2}\Omega_{n,0}x_0^{-n}n(n-3).
\label{eq:23}
\end{equation}
Hence, if $n>3$ we obtain 
$\left(\frac{\partial^2 V}{\partial x^2}\right)\left(x_0\right)$ positive 
which corresponds to the centres in the phase plane. The presence of centres 
on the phase portraits means that all bouncing models are oscillating. Let us 
note, however that the corresponding systems are structurally unstable because 
of the presence of nonhyperbolic critical points on the phase portraits. 
Physically it means that the small perturbation of right-hand sides of the 
system under consideration disturbs a qualitative structure of the orbits 
(i.e. a phase portrait). On Fig.~\ref{fig:1} and \ref{eq:2} the phase 
portraits and diagrams of the potential function are presented for the 
standard and generalized bouncing model. Fig. \ref{fig:1} describes all 
special cases listed in Table~\ref{tab:1}. The classically forbidden region 
for $a<a_{0}$ is shaded. The evolution of the model is represented in the 
configuration space by a Hamiltonian level
\[
\mathcal{H} = E = \frac{1}{2}\Omega_{k,0}.
\]
The trajectory of the flat model separates the regions occupied by both closed 
and open models. The decreasing of the potential function with respect to the
scale factor determines the domain of phase space occupied by accelerating
trajectories. The bounce is the intersection point of trajectory with
the axis $a$. Note that around the bounce we have acceleration. On the phase
plane of Hamiltonian dynamical systems only centres and saddle points are
admissible. The centres are structurally unstable while saddles are
structurally stable. Because the centre appears in the phase portraits of
standard and generalized bouncing models, both models are structurally
unstable. The critical points represent the static universes. Generalized
bouncing model has two disjoint acceleration regions. The first is due to
bouncing term while the second is forced by the cosmological constant term.

\begin{table}
\caption{Some special cases of bouncing models}
\label{tab:1}
\begin{tabular}{c|c}
\hline
model & dynamical equations (first integral) \\ \hline
FRW model dust filled
& $\dot{x} = yx^3$\\
universe with global rotation \cite{Godlowski:2003hf}
& $\dot{y}=\frac{1}{2}\left(-\Omega_{m,0}x^{-2}+2|\Omega_{\omega,0}|x^{-3}\right)
x^3$ \\
or brane models with dark radiation \cite{Vishwakarma:2003}
& $\frac{y^2}{2}=
\frac{1}{2}\left(-\Omega_{m,0}x^{-1}-|\Omega_{\omega,0}|x^{-2}+\Omega_{k,0}\right)
x^6$\\
\hline
FRW dust filled
& $\dot{x} = yx^5$\\
universe with spinning fluid  \cite{Szydlowski:2004c}
& $\dot{y}=\frac{1}{2}\left(-\Omega_{m,0}x^{-2}+4|\Omega_{s,0}|x^{-5}\right)
x^5$ \\
or a class of MAG models \cite{Krawiec:2005jj}
& $\frac{y^2}{2}=
\frac{1}{2}\left(-\Omega_{m,0}x^{-1}-|\Omega_{s,0}|x^{-4}+\Omega_{k,0}\right)
x^{10}$\\
\hline
Stephani models
& $\dot{x}=y $\\
filled by perfect fluid
& $\dot{y}=\frac{1}{2}\left(-\Omega_{\gamma,0}(1+3\gamma)x^{-2-3\gamma}
+\delta\Omega_{k,0}x^{\delta-1}\right) $\\
$p=\gamma \rho$ \cite{Godlowski:2004c}
& $\frac{y^2}{2} = \frac{1}{2}\left(\Omega_{\gamma,0}x^{-1-3\gamma}
+\Omega_{k,0}x^{\delta}\right)$\\
\hline
\end{tabular}
\end{table}

\begin{figure}
\includegraphics[width=0.7\textwidth]{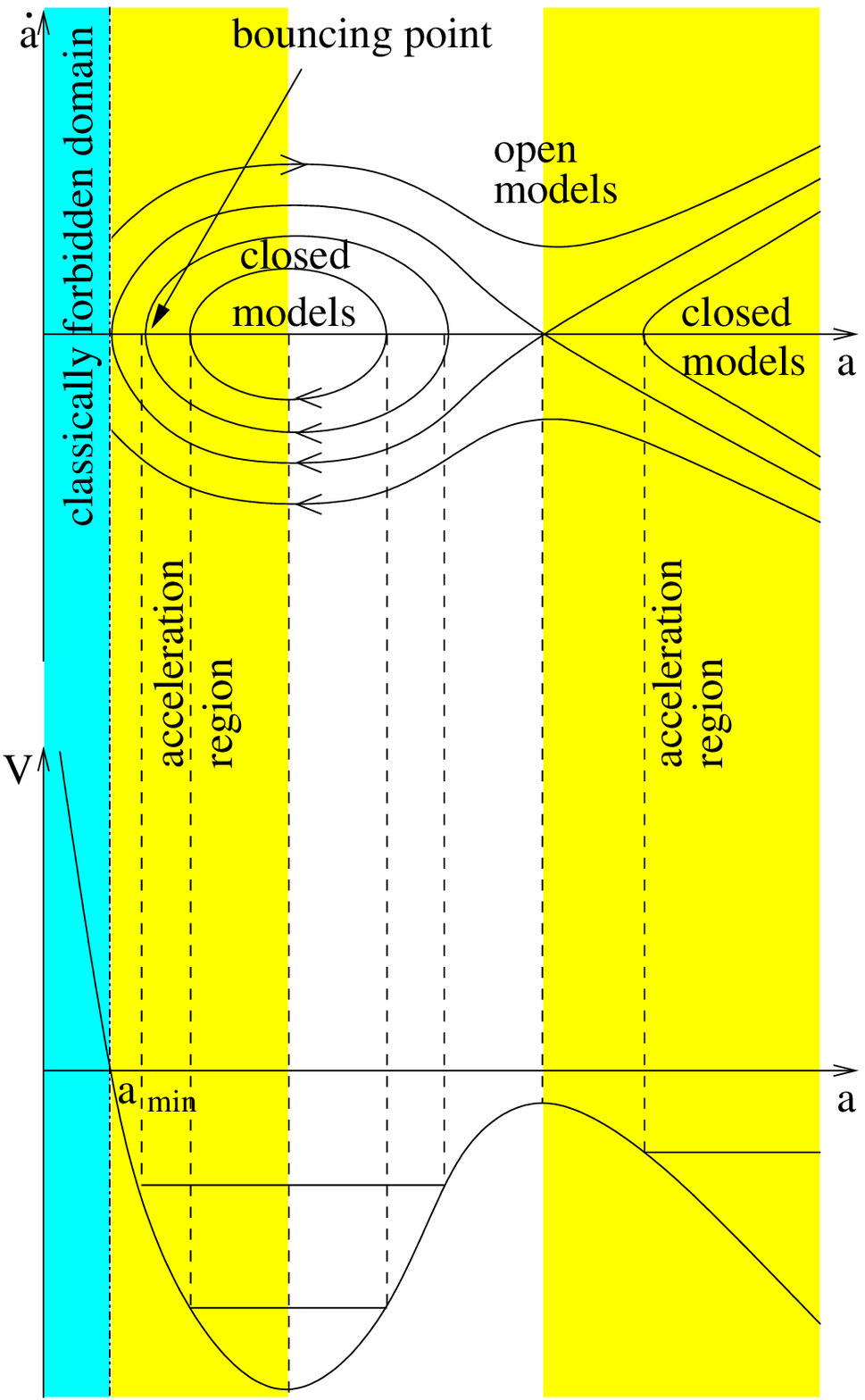}
\caption{The phase portrait and the diagram of the potential function for
$\Lambda$BCDM model. The minimum (maximum) of the potential
function corresponds to a centre (saddle) on the phase plane. The system
is structurally unstable because of the presence of nonhyperbolic critical
point (a centre).}
\label{fig:2}
\end{figure}

The full knowledge of the dynamics required its analysis at infinity, i.e. at 
the circle at infinity $x^2+y^2=\infty$. The standard procedure is to use 
projective maps on the plane and then analyze the system in a standard way. 
One can find the critical points at infinity as an a intersections of 
trajectory of a flat model with circle at infinity, i.e.
$\{ (x,y) \colon \Omega_{k,0}=0 \}$ and 
$\{ z=1/x, u=y/x, \quad \text{and} \quad x=\infty, v=1/y, w=x/y, y=\infty \}$ 
-- the trajectory of the flat model $\frac{y^2}{2}=-V(x)$ with a circle at 
infinity.
 
Some special cases of the bouncing systems contain Table \ref{tab:1}. Because 
we are dealing with autonomous dynamical systems their phase portrait is always 
given modulo diffeomorphism or equivalent modulo any time reparameterization 
following the rule: $\tau \to \eta \colon d\tau=f(x)d\eta$,
where $f(x)$ is diffeomorphism, $x$ is point which belong to phase plane.
 
For analysis of bouncing models in term of dynamical systems, it is useful to
reparameterized original time variable in order to obtain nondegenerate
critical   points at infinity. Then we obtain $\frac{d\tau}{x^{\beta/2}}$ and
$\frac{y^2}{2}=\frac{1}{2}(\Omega_{m,0}x^{2-m+\beta}-\Omega_{n,0}x^{2-n+\beta})$
is now representing the trajectory of the flat model, $\beta$ should be chosen
in the suitable way to regularize critical points. Let $\beta=n-2$ then as
$x$ goes to infinity then $y$ also goes to infinity. Only the sign of the 
parameter $m$ (if $m<0$) decides whether the future of the system is the type 
of a big rip singularity. If $m=0$ then the case of cosmological constant can 
be recovered.
 
It is convenient to regularize the system by multiplication both sides of
the system $x^3$ in the first case and $x^4$ in the second one respectively.
It is equivalent to reparameterized time variable following rule
$\tau \to \eta \colon \frac{d\tau}{x^{\beta}}=d\eta$, where $\beta=3$ and (5)
for the system from Table~\ref{tab:1}. For both systems from the table we 
have an additional term in the generalized FRW equation. In the first case the 
effects of global rotation produce contribution corresponding to the negative 
energy scaling like radiation. The same contribution appears in the brane 
models on the charged brane. It is known as the dark radiation 
\cite{Dabrowski:2004}. Please note, that analogous term appeared if we include 
the Casimir effect coming for example from quantum effects of massless scalar 
fields \cite{Szydlowski:1987vr,Szydlowski:1989b,Coule:2004qf}. In the second 
case (Table~\ref{tab:1}) it is the model with spinning dust fluid. It can be 
also recovered as a class of MAG models \cite{Puetzfeld:2004,Krawiec:2005jj}.
 
In both cases we can find centre at finite domain and periodic orbits.
At infinity we have unstable and stable nodes at $x=+\infty, y=\mp \infty$.
The trajectory of the flat model separates the regions occupied by closed
and open models. All models have  bounce but some from them are oscillating
models without the initial and final singularities. For our future
investigations of observational constraints on bouncing models it is convenient
to derive crucial formulae for $H(z)$ where $z$ is redshift 
$z \colon 1+z=x^{-1}$. We obtain from (\ref{eq:9}) that
$H^2=H^2_0\left[\Omega_{m,0}(1+z)^{m}-\Omega_{n,0}(1+z)^{n}\right]$.
It is useful to represent it in the corresponding bouncing parameters.
 
For this aim we find $x_b$ corresponding to the bounce and value of redshift
which identify this moment during the evolution:
$x_b=\left(\frac{\Omega_{n,0}}{\Omega_{m,0}}\right)^{\frac{1}{n-m}}$,
$z_b=-1+\left(\frac{\Omega_{m,0}}{\Omega_{n,0}}\right)^{\frac{1}{n-m}}$.
Finally we obtain independent model parameters characterizing its role in
evolution (modulo present value of $H_0$), namely
\begin{equation}
H=H_0 \sqrt{\frac{\left(1+z_b\right)^{n-m}}{\left(1+z_b\right)^{n-m}-1}}
\left(1+z\right)^{m/2} \sqrt{1+ \left(\frac{z+1}{z_b+1}\right)^{n-m}}.
\label{eq:24}
\end{equation}
 
If $\Omega_{m,0}$ is fixed, for example from independent galaxy observations
then the evolutionary scenario is parameterized by single $n$ parameter
\begin{equation}
H=H_0 \sqrt{\Omega_{3,0}}\left(1+z\right)^{3/2}
\sqrt{1+ \left(1-\frac{1}{\Omega_{3,0}}\right)\left(1+z\right)^n},
\label{eq:25}
\end{equation}
where we put $\Omega_{m,0}=\Omega_{3,0}$, i.e. dust filled universe.
 
In the case of generalized bouncing models the potential function takes
the following form
\[
V(x) = \frac{1}{2} \Omega_{n,0}x^{2-n} - \frac{1}{2} \Omega_{m,0}
x^{2-m} - \frac{1}{2}\Omega_{\Lambda,0} x^{2}.
\]

It means that if only $n,m>0$ then we obtain the de Sitter solution as a 
global attractor in the future. In the opposite case if $m>0$ the big-rip 
singularities are generic future of the model. Note that in the class of 
generalized bouncing models only trajectories around point $(x_{0},0)$ 
represent oscillating models without a singularity and there is admissible 
large class of closed, open and flat models which evolve to infinity.

It is interesting that the characteristic bounce can be defined in terms of
geometry of potential function only. By bouncing cosmology we can understand 
all cosmological models for which the potential function has at some point 
a minimum.

\section{Bouncing model and distant supernovae observations.}
 
In this section we confront the bouncing cosmological models with observations
of distant SNIa. These observations in the framework of the FRW model indicate
that present acceleration of our Universe is due to an unknown form of matter
with negative pressure called dark energy \cite{Perlmutter:1999}. Apart from
the cosmological constant there are also other candidates for dark energy
which were tested from SNIa observations 
\cite{Alam:2003rw,Gorini:2004b,Biesiada:2005}.
We use the SNIa data to test the acceleration in the bouncing models.
Moreover these models are attractive because they have no horizon and initial
singularity, and they yield an explanation of structures which originated in
the quantum epoch \cite{Coule:2004qf}.
 
We consider the flat FRW model since there is a very strong evidence that the 
Universe is flat in the light of recent WMAP data \cite{Bennett:2003bz}. We 
confront the two ``bouncing'' models (with and without extra $\Lambda$ fluid) 
with SNIa data. For this purpose we calculate the luminosity distance in 
a standard way 
\begin{equation}
d_{L}(z)=(1+z)\int_{0}^{z} \frac{d\bar{z}}{H(\bar{z})}
\label{eq:104}
\end{equation}
To proceed with fitting models to SNIa data we need the magnitude-redshift
relation
\begin{equation}
m(z,\mathcal{M},\Omega_{m,0},\Omega_{\Lambda,0},n,m)-M
=\mathcal{M}+5\log_{10}D_{L}(z,\Omega_{m,0},\Omega_{\Lambda,0},m,n)
\label{eq:105}
\end{equation}
where $M$ being the absolute magnitude of SNIa and
\begin{equation}
D_{L}(z,\Omega_{m,0},\Omega_{\Lambda,0},m,n)
=H_{0}d_{L}(z,H_{0},\Omega_{m,0},\Omega_{\Lambda,0},m,n)
\label{eq:105a}
\end{equation}
is the luminosity distance with $H_{0}$ factored out, so that marginalization
over the parameter $\mathcal{M}$
\begin{equation}
\mathcal{M}=-5\log _{10}H_{0}+25
\label{eq:106}
\end{equation}
reads actually marginalization over $H_{0}$.
 
The parameter $\mathcal{M}$ is actually determined from the
low-redshift part of the Hubble diagram which should be linear
in all realistic cosmologies. It lead to value of $H_0 \simeq 65$ km/s Mpc
\cite{Perlmutter:1999,Riess:1998,Riess:2004}, i.e., $\mathcal{M} \simeq 15.955$.
In further analysis we estimate the models with this value of $\mathcal{M}$
and without any prior assumption on $H_0$.
 
Then we can obtain the best fit model minimizing the function $\chi^{2}$
\begin{equation}
\chi^{2}=\sum_{i}\frac{(\mu_{i}^{theor}-\mu_{i}^{obs})^{2}}{\sigma_{i}^{2}}
\label{eq:107}
\end{equation}
where the sum is over the SNIa sample and $\sigma_{i}$ denote the (full)
statistical error of magnitude determination and $\mu_i=m_i-M_i$.
 
Because the best-fit values alone are not sufficient, the statistical analysis
is supplemented with the confidence levels for the parameters. We performed
the estimation of model parameters using the minimization procedure, based on
the maximum likelihood method. We assume that supernovae measurements came
with uncorrelated Gaussian errors and the likelihood function $\mathcal{L}$
could be determined from the chi-square statistic
$\mathcal{L}\propto \exp(-\chi^{2}/2)$ \cite{Riess:1998}.
 
The first published large samples of SNIa appeared at the end of the 90s
\cite{Perlmutter:1999,Riess:1998}. Later other data sets have been made either
by correcting errors or by adding new supernovae. The latest compilation
of SNIa was prepared by Riess et al. \cite{Riess:2004} and became de facto
a standard data set. It should be noted that this compilation encloses
the largest number of high-redshift $z>1.25$ objects in compare to older
compilations. From this compilation we take the ``Silver'' sample which
contains all 186 SNIa, and the restricted ``Gold'' sample of 157 SNIa (with
higher quality of the spectroscopic and photometric records).
 
In order to test a cosmological model we calculate the best fit with minimum 
$\chi^2$ as well as estimate the model parameters using the maximum likelihood 
method \cite{Riess:1998}. For both statistical methods we take the parameters 
$m$ and $n$ in the interval $[0,10]$, $n>m$. We test separately the models 
with and without the cosmological constant term. We also assume priors about 
$\Omega_{\text{m},0}$ and we estimate it or take $\Omega_{\text{m},0}= 0.3$ 
(baryonic plus dark matter in galactic halos) \cite{Ratra:1988}.

The results of two fitting procedures performed on the ``Gold'' sample for 
the cosmological bouncing models with different prior assumptions are 
presented in (Table~\ref{tab:2} and \ref{tab:3}). These tables refer both 
to the $\chi^2$ (best fit) and results from marginalized probability of 
density functions.

\begin{table}
\caption{Results of the statistical analysis of the bouncing model without
dust (BM) and bouncing cold dark matter model (BCDM) obtained
for SNIa data from the best fit with minimum $\chi^2$ (denoted as
BF) and from the likelihood method (denoted as L). The case of
a fixed value of the parameter $\mathcal{M}$ is denoted as F.
If in BF method we obtain $\Omega_{\text{n},0}=0$ than $n$ could be
taken arbitrary (marked as $A$).}
\label{tab:2}
\begin{tabular}{c|rrrrrrr}
\hline \hline
model & $\Omega_{\mathrm{m},0}$&  $m$ & $\Omega_{\mathrm{n},0}$&
$n$ &  $\mathcal{M}$ & $\chi^2$& method \\
\hline
BM model        &  1.00 & 1.4  & 0.00 & A    & 15.975 &181.6 &  BF  \\
                &  1.00 & 1.5  & 0.00 & 1.7  & 15.975 & ---  &  L   \\
                &  1.54 & 1.4  & 0.54 & 1.5  &F15.955 &182.3 &  BF  \\
                &  1.00 & 1.4  & 0.00 & 1.6  &F15.955 & ---  &  L   \\
\hline
BCDM model              &  1.86 &---  & 0.86 & 3.7  & 16.085 &217.4 &  BF  \\
(dust matter $m=3$)     &  1.86 &---  & 0.86 & 3.7  & 16.095 & ---  &  L   \\
                        &  1.86 &---  & 0.86 & 3.7  &F15.955 &273.7 &  BF  \\
                        &  1.86 &---  & 0.86 & 3.7  &F15.955 & ---  &  L   \\
\hline
\end{tabular}
\end{table}

\begin{table}
\caption{The results of statistical analysis of BCDM models ($m=3$)
obtained for SNIa data from the best fit with minimum
$\chi^2$ (denoted as BF) and from the likelihood method (denoted as L).
The case of a fixed value of ${\cal M}$ is denoted as F.}
\label{tab:3}
\begin{tabular}{c|rrrrr}
\hline \hline
model & $\Omega_{\mathrm{m},0}$& $\Omega_{\mathrm{n},0}$&
$\mathcal{M}$ & $\chi^2$& method \\
\hline
BCDM model          &  1.50 & 0.50 & 16.105 &226.6 &  BF  \\
with $n=4$          &  1.50 & 0.50 & 16.095 & ---  &  L   \\
                    &  1.50 & 0.50 &F15.955 &296.4 &  BF  \\
                    &  1.50 & 0.50 &F15.955 & ---  &  L   \\
\hline
BCDM model          &  1.03 & 0.03 & 16.175 &291.2 &  BF  \\
with $n=6$          &  1.03 & 0.03 & 16.175 & ---  &  L   \\
                    &  1.03 & 0.03 &F15.955 &443.4 &  BF  \\
                    &  1.03 & 0.03 &F15.955 & ---  &  L   \\
\hline
\end{tabular}
\end{table}

At first we analyzed bouncing model without any priors for $m$ parameter (BM).
We obtain the value $\chi^2=181.6$ what means that this model is acceptable
on the $2\sigma$ level with degree of freedom $\text{df}=153$. However,
the estimated value of $m=1.4$ in the model is unrealistic because the dust
matter is present in the universe ($m=3$). With the prior $m=3$ we obtain
$\chi^2=217.4$ with the value of the parameter $n=3.7$. For the more realistic
model with $m=3$ and $n=4$ (because of the presence of radiation matter in
the Universe) (Table~\ref{tab:3}) we obtain $\chi^2=226.6$. While the
bouncing model with dust (BCDM) is better fitted than the Einstein-De Sitter
model it is rejected at least on the $4\sigma$ level. With priors
$\mathcal{M} \simeq 15.955$ the model is rejected on the $8\sigma$ level.
 
\begin{figure}
\includegraphics[width=0.8\textwidth]{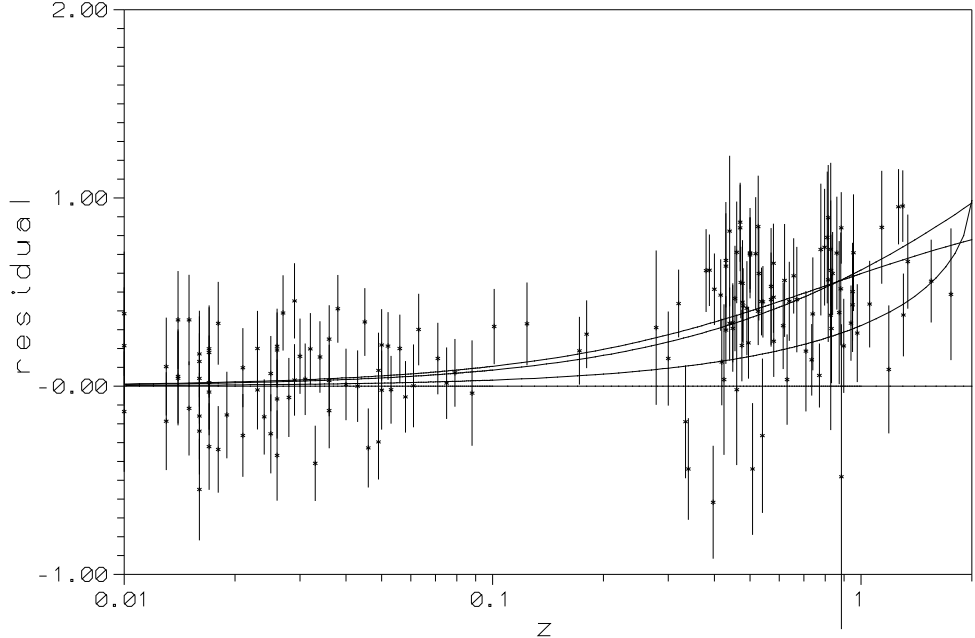}
\caption{Residuals (in mag) between the Einstein-de Sitter model and
the Einstein-de Sitter itself (zero line), the $\Lambda$CDM flat model
(upper curve) the best-fitted BM model, (upper-middle curve)
and the best-fitted BCDM model with $m=3$ (lower-middle curve)
(with assumed {\cal M}=15.955).}
\label{fig:3}
\end{figure}

In Fig.~\ref{fig:3} we present of residuals plots of the $m$-$z$ relation
for considered models with respect to the Einstein-de Sitter (CDM) model.
Apart the CDM model (the zero line) the three models $\Lambda$CDM, BM and
BCDM are shown. The diagrams for bouncing models intersect the $\Lambda$CDM
diagram in such a way that the supernovae on intermediate distances are
brighter then expected in the $\Lambda$CDM model, while very high redshift
supernovae should be fainter then they are expected in the $\Lambda$CDM model.
Note that this effects are more stronger for the BCDM model than for the BM
model.

Similarly we analyze the bouncing models with the additional parameter---the
cosmological constant. We fixed value of $m=3$ (dust matter) and it is called
the extended bouncing model ($\Lambda$BCDM). This model with $\text{df}=153$
is statistically admissible on the $2\sigma$ level (Table~\ref{tab:4}),
but we obtain $\Omega_{n,0}=0$ (no bounce term) as a most probable value.
This result is similar also for models with fixed values $n=4$ and $n=6$
(Table~\ref{tab:5}), as well as it is independent from the assumption
on $\Omega_{\text{m},0}$. In this way $\Lambda$BCDM reduces to ``classical''
$\Lambda$CDM model.

\begin{table}
\caption{Results of the statistical analysis of the extended bouncing models
($m=3$), obtained for SNIa data from the best fit with minimum $\chi^2$
(denoted as BF) and from the likelihood method (denoted as L). The case of a
fixed value of parameter $\Omega_{m,0}$ is denoted as F.
If in BF method we obtain $\Omega_{\text{n},0}=0$ than $n$ could be
taken arbitrary (marked as $A$).}
\label{tab:4}
\begin{tabular}{c|rrrrrrr}
\hline \hline
model & $\Omega_{\mathrm{m},0}$&  $\Omega_{\mathrm{n},0}$&
$n$ & $\Omega_{\Lambda,0}$& $\mathcal{M}$ & $\chi^2$& method \\
\hline
$\Lambda$BCDM model &  0.31 & 0.00 &  A.  & 0.69 & 15.955 &175.9 &  BF  \\
                    &  0.34 & 0.00 &  3.0 & 0.67 & 15.965 & ---  &  L   \\
                    & F0.30 & 0.00 &  A   & 0.70 & 15.955 &175.9 &  BF  \\
                    & F0.30 & 0.00 &  3.0 & 0.70 & 15.945 & ---  &  L   \\
\hline
$\Lambda$BCDM model    &  0.31 & 0.00 &  A   & 0.69 & --- &175.9 &  BF  \\
with ${\cal M}=15.955$ &  0.31 & 0.00 &  3.0 & 0.68 & --- & ---  &  L   \\
                       & F0.30 & 0.00 &  A   & 0.70 & --- &175.9 &  BF  \\
                       & F0.30 & 0.00 &  3.0 & 0.70 & --- & ---  &  L   \\
\hline
\end{tabular}
\end{table}

\begin{table}
\noindent
\caption{Results of comparison of $\Lambda$CDM model with the extended
bouncing models ($m=3$) with fixed values $n=4$ and $n=6$.
The result of statistical analysis for SNIa data from the best fit with
minimum $\chi^2$ (denoted as BF) and from the likelihood method (denoted
as L). The case of a fixed value of $\Omega_{m,0}$ is denoted as F.}
\label{tab:5}
\begin{tabular}{c|rrrrrr}
\hline \hline
model & $\Omega_{m,0}$&  $\Omega_{n,0}$&
$\Omega_{\Lambda,0}$&  $\mathcal{M}$ & $\chi^2$& method \\
\hline
$\Lambda$CDM model   &  0.31 & --- & 0.69 & 15.955 &175.9 &  BF  \\
                     &  0.34 & --- & 0.67 & 15.965 & ---  &  L   \\
                     & F0.30 & --- & 0.70 & 15.955 &175.9 &  BF  \\
                     & F0.30 & --- & 0.70 & 15.945 & ---  &  L   \\
\hline
$\Lambda$BCDM model  &  0.31 & 0.00 & 0.69 & 15.955 &175.9 &  BF  \\
with $n=4$           &  0.37 & 0.00 & 0.65 & 15.965 & ---  &  L   \\
                     & F0.30 & 0.00 & 0.70 & 15.955 &175.9 &  BF  \\
                     & F0.30 & 0.00 & 0.70 & 15.945 & ---  &  L   \\
\hline
$\Lambda$BCDM model  &  0.31 & 0.00 & 0.69 & 15.955 &175.9 &  BF  \\
with $n=6$           &  0.34 & 0.00 & 0.66 & 15.965 & ---  &  L   \\
                     & F0.30 & 0.00 & 0.70 & 15.955 &175.9 &  BF  \\
                     & F0.30 & 0.00 & 0.70 & 15.945 & ---  &  L   \\
\hline
\end{tabular}
\end{table}
 
\begin{figure}
\includegraphics[width=0.8\textwidth]{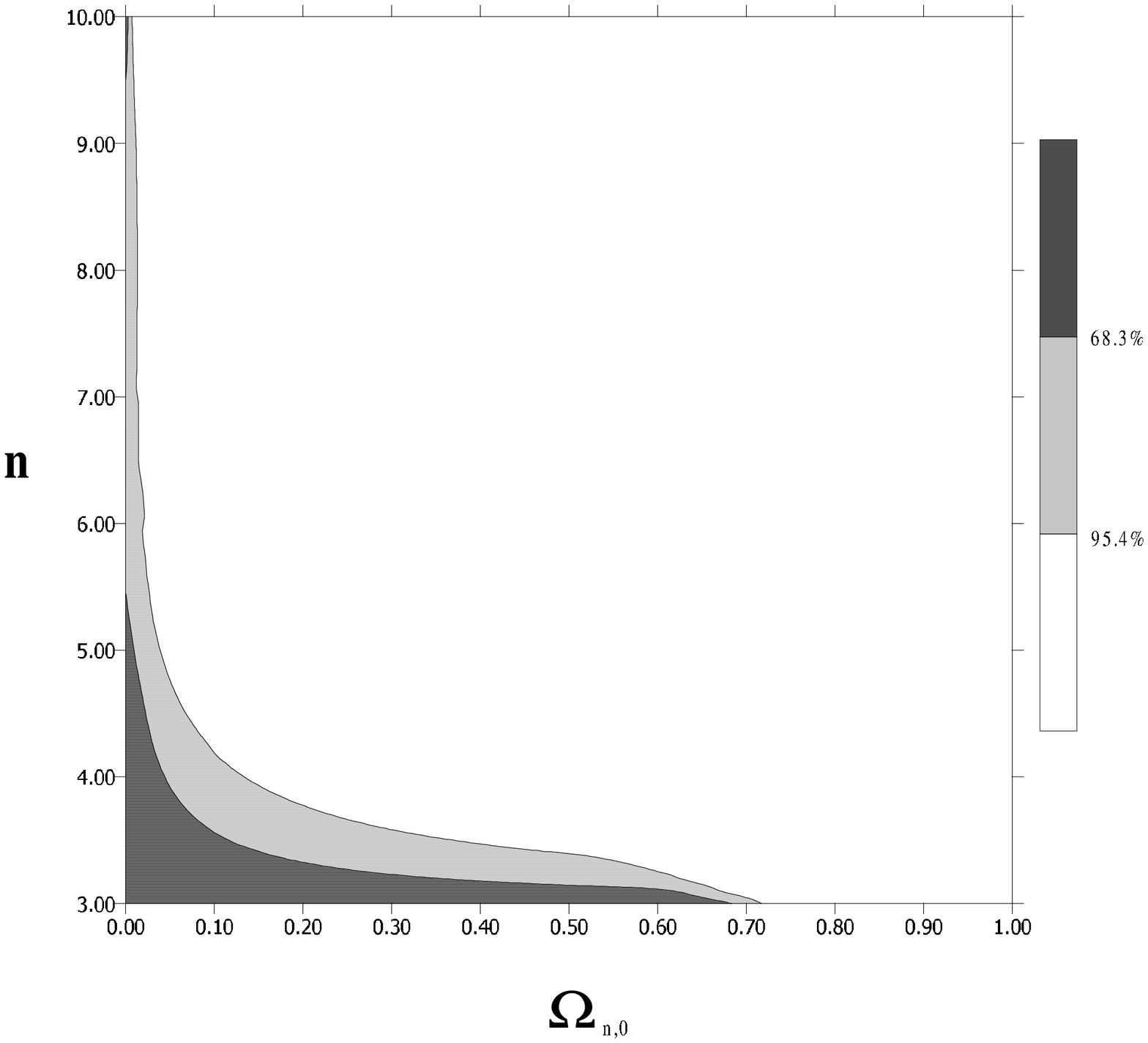}
\caption{For the extended bouncing model with
${\cal M}=15.955$) there are shown the confidence levels on the plane
$(\Omega_{n,0},n)$ minimized over parameter $\Omega_{m,0}$.}
\label{fig:4}
\end{figure}

\begin{figure}
\includegraphics[width=0.8\textwidth]{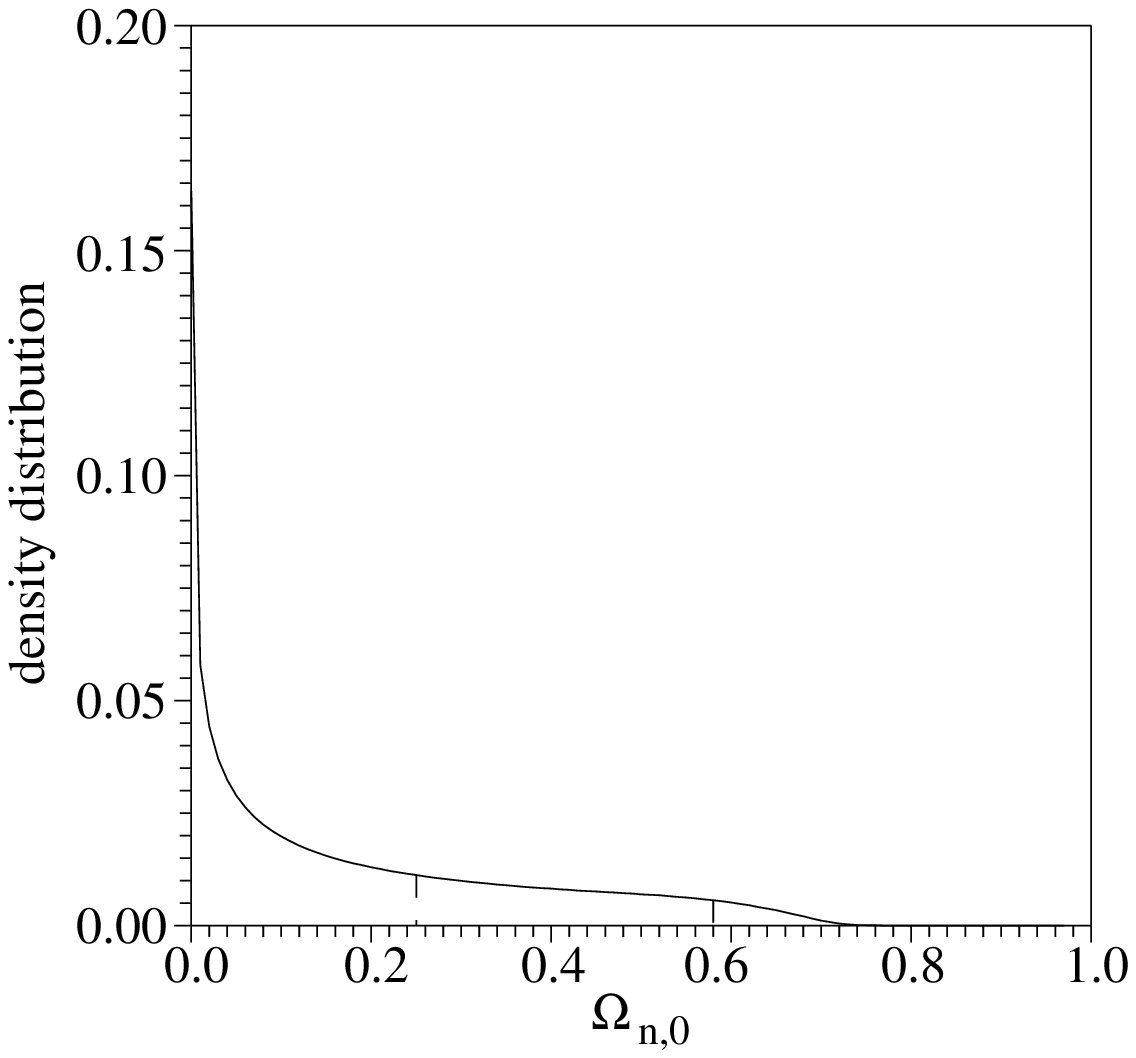}
\caption{Extended Bouncing model with {\cal M}=15.955. The density
distribution for $\Omega_{n,0}$. Confidence level $68.3\%$ and $95.4\%$ are
also marked on the figure.}
\label{fig:5}
\end{figure}

\begin{figure}
\includegraphics[width=0.8\textwidth]{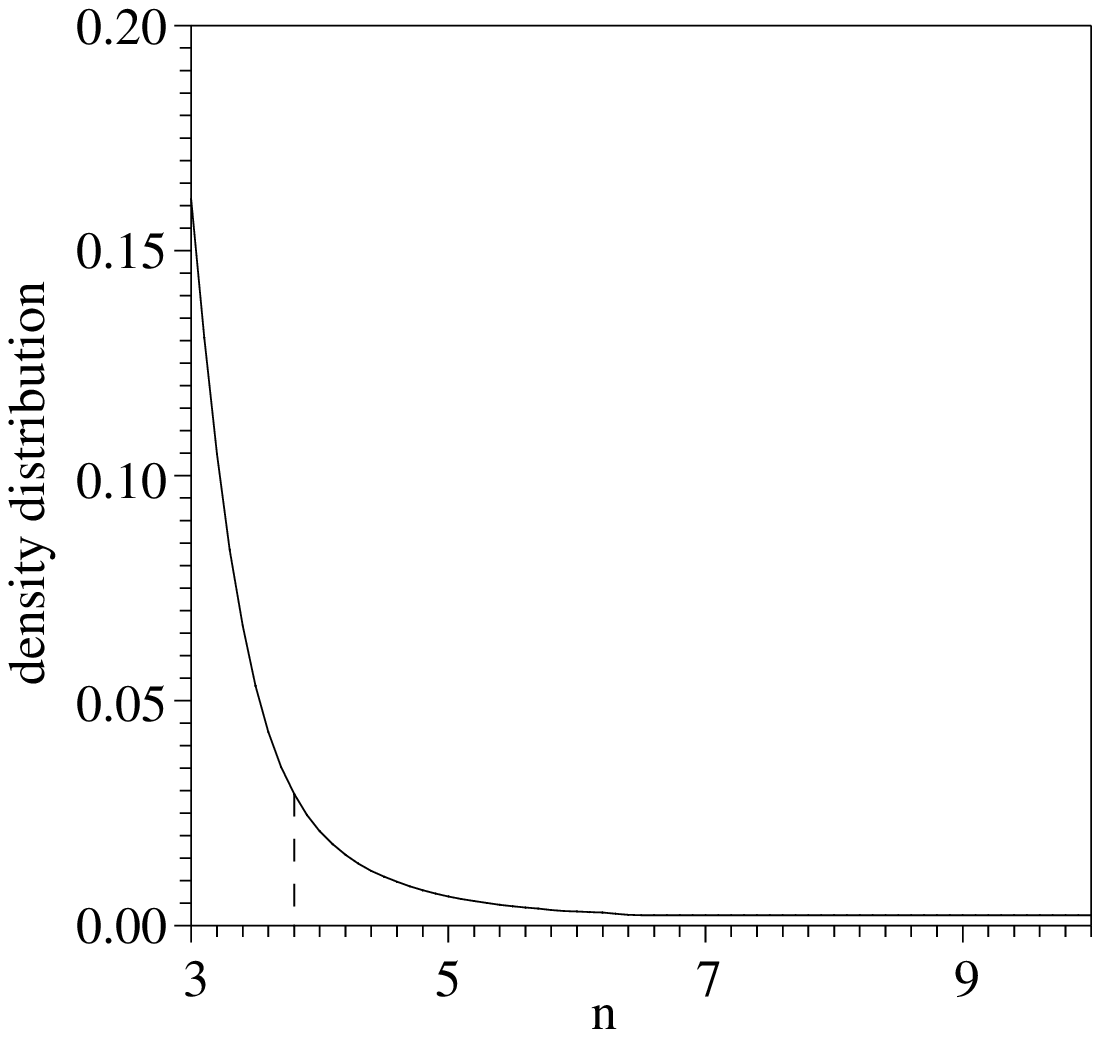}
\caption{Extended Bouncing model with {\cal M}=15.955. The density
distribution for $n$. Confidence level $68.3\%$ and $95.4\%$ are
also marked on the figure.}
\label{fig:6}
\end{figure}

The confidence levels in the $(\Omega_{n,0},n)$ plane are presented in
Fig. \ref{fig:4}. In order to complete the picture we have also derived
one-dimensional probability distribution functions (PDF) for $\Omega_{n,0}$
(Fig. \ref{fig:5}) and $n$ (Fig. \ref{fig:6}) obtained from the joint
marginalization over remaining model parameters. The maximum value of such
a PDF informs us about the most probable value of $\Omega_{n,0}$, supported
by supernovae data within the extended bouncing dust model.

From the PDFs the most probable value of $\Omega_{n,0}$ is also equal $0$,
however non-zero value of $\Omega_{n,0}$ cannot be excluded.
In this way, it is crucial to determine which combination of parameters
give the preferred fit to data. This is the statistical problem of
model selection \cite{Liddle:2004nh}. The problem is usually the elimination of
parameters which play insufficient role in improving the fit data available.
Important role in this area plays especially the Akaike
information criterion (AIC) \cite{Akaike:1974}. This criterion is defined as
\begin{equation}
\label{eq:244}
AIC=-2\ln{\mathcal{L}}+2k
\end{equation}
where $\mathcal{L}$ is the maximum likelihood and $k$ is the number of the
parameter of the model. The best model is the model which minimizes the
AIC. The AIC for the models under consideration is presented in the
(Table~\ref{tab:6}). It is clear that model which minimizing AIC is
$\Lambda$CDM. Therefore it is no reason to introduce a model with bouncing
terms and such model should be ruled out by Occam's razor. Because
the extended bouncing dust model is statistically admissible from SNIa data
it can be reconsidered only if the firm theoretical reason appears.
Only this situation can justify consideration of the model with a small, but 
non-zero bouncing term.
 
\begin{figure}
\includegraphics[width=0.8\textwidth]{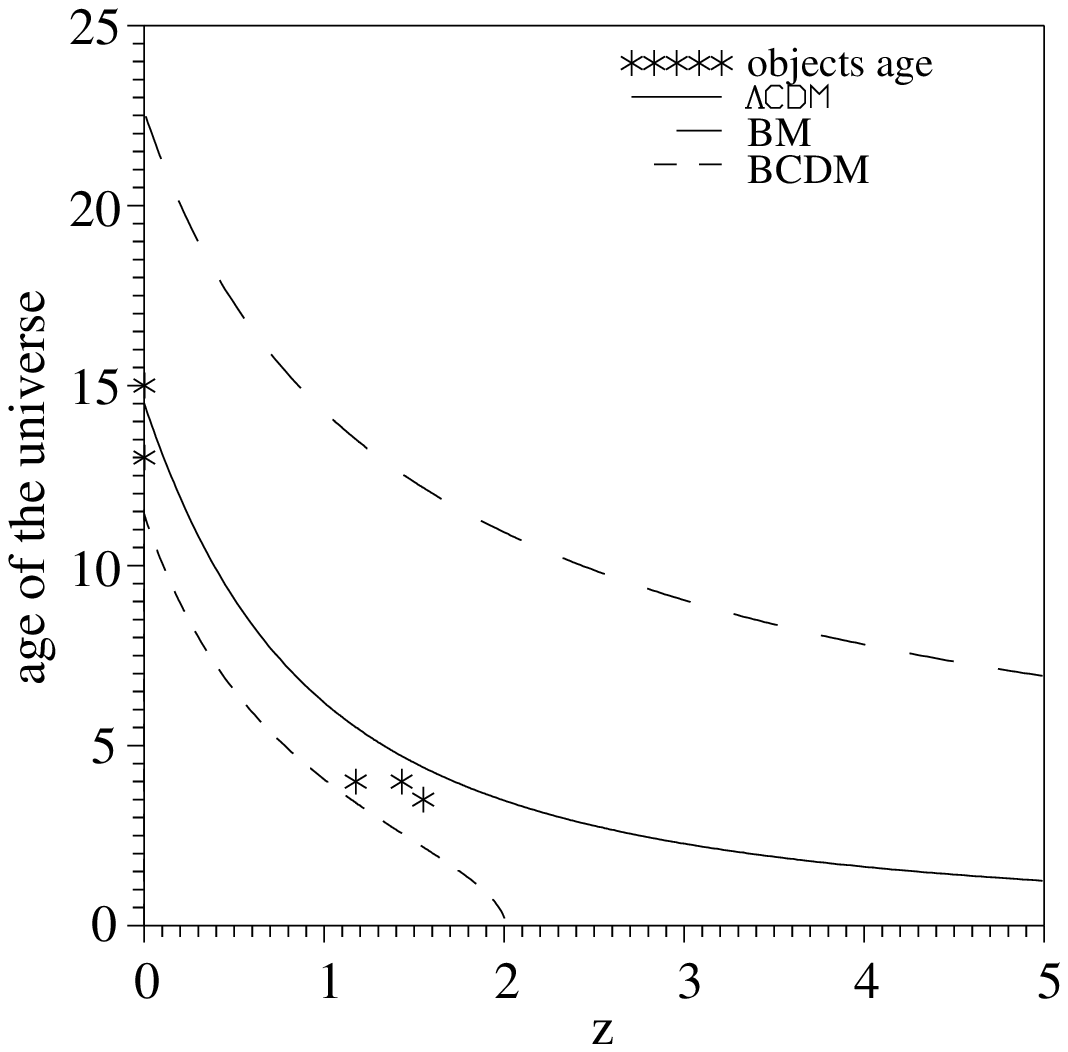}
\caption{The age of the universe on particular $z$ for three class of models:
$\Lambda$CDM (middle curve), bouncing model BM (upper curve) and bouncing model
with dust matter BCDM (lower curve). We marked age of 4 extragalactic object
by star (Table~\ref{tab:7}).}
\label{fig:7}
\end{figure}

The existence of the oldest high-redshift extragalactic (OHReG) objects could be
used as a test of the cosmological models (Table~\ref{tab:7}). The globular
cluster analysis indicated that the age of the Universe is $13.4$ Gyr
\cite{Chaboyer:2003}. We demonstrate that the age of OHReG objects restricts
the model parameter. As a criterion we take that the age of the Universe in
a given redshift should be bigger than, at least equal, to the age of its
oldest objects. With the assumption of $H_{0} = 65$ km/s MPc, the age of the
universe on particular $z$ for three class of models is calculated
(Fig. \ref{fig:7}). This test admits the $\Lambda$CDM model with
$\Omega_{\text{m},0}=0.3$. In this model, the age of the Universe is
$14.496$ Gyr. The BM model seems to be allowed from this test, however, that
model predicts much longer age of the Universe (more than 20 Gyr) then
$\Lambda$CDM. In turn the BCDM model must be rejected because its age is
$11.5$ Gyr.

\begin{table}
\noindent
\caption{The Akaike information criterion (AIC) for models under consideration:
Einstein-de Sitter model (CDM), $\Lambda$CDM model ($\Lambda$CDM),
bouncing model (BM), bouncing model with dust $m=3$ (BCDM)
and extended bouncing model with dust $m=3$ ($\Lambda$BCDM).}
\label{tab:6}
\begin{tabular}{c|cr}
\hline \hline
model & no. of parameters &  AIC \\
\hline
CDM             & 1  & 325.5   \\
$\Lambda$CDM    & 2  & 179.9   \\
BM              & 4  & 189.6   \\
BCDM             & 3  & 223.4   \\
BCDM with $n=4$  & 2  & 230.6   \\
BCDM with $n=6$  & 2  & 295.2   \\
$\Lambda$BCDM            & 4  & 183.9   \\
$\Lambda$BCDM with $n=4$ & 3  & 181.9   \\
$\Lambda$BCDM with $n=6$ & 3  & 181.9   \\
\hline
\end{tabular}
\end{table}

\begin{table}
\noindent
\caption{The age of extragalactic objects.}
\label{tab:7}
\vspace{0.2cm}
\begin{tabular}{@{}p{1.5cm}rrrr}
\hline \hline
No & object  & $z$ & age in Gys \\
\hline
1   & globular cluster &$ 0.$    & $13 - 15$\\
2   & 3C65 quasar      &$ 1.175$ & $4.0    $\\
3   & LBDS 53W069      &$ 1.43 $ & $4.0    $\\
4   & LBDS 53W091      &$ 1.55 $ & $3.5    $\\
\hline
\end{tabular}
\end{table}

\section{CMB peaks in the extended bouncing model}

Acoustic oscillations in the primeval plasma during the last scattering give
rise to the temperature map of cosmic microwave background (CMB). Peaks in
the power spectrum correspond to maximum density of the wave. In the Legendre
multipole space these peaks correspond to the angle subtended by the sound
horizon at the last scattering. Further peaks answer to higher harmonics of
the principal oscillations.
 
The locations of these peaks depend on the variations in the model
parameters. Therefore, they can be used to constrain the parameters of
cosmological models.
 
The acoustic scale $\ell_{A}$ which puts the locations of the peaks is defined
as
\begin{equation}
\label{eq:44}
\ell_{A} = \pi \frac{\int_{0}^{z_{\text{dec}}} \frac{dz'}{H(z')}}
{\int_{z_{\text{dec}}}^{\infty} c_{s} \frac{dz'}{H(z')}}
\end{equation}
where
\begin{equation}
\label{eq:45}
H(z) = H_{0} \sqrt{\Omega_{\text{m},0}(1+z)^3 + \Omega_{\text{r},0}(1+z)^4-
\Omega_{n,0}(1+z)^n+\Omega_{\Lambda,0}}
\end{equation}
and $c_{\text{s}}$ is the speed of sound in the plasma given by
\begin{equation}
\label{eq:46}
c_{\text{s}}^{2} \equiv \frac{dp_{\text{eff}}}{d\rho_{\text{eff}}} =
\frac{\frac{4}{3} \Omega_{\gamma,0}(1+z) -\frac{n-3}{3} n \Omega_{n,0}(1+z)^{n-3}}
{3 \Omega_{\text{b},0} + 4 \Omega_{\gamma,0}(1+z) - n\Omega_{n,0}(1+z)^{n-3}}.
\end{equation}
 
The properties of the bouncing term $\Omega_{n,0}$ are unknown. In particular,
we do not know whether it influences the sound velocity. But we assume that
sound can propagate in it as well as in baryonic matter and photons.
Let us note that with the lack of the bouncing term (i.e. $\Omega_{n,0}=0$)
and/or when sound does not propagate in the bouncing fluid,
we obtain the standard formula for $c_{\text{s}}^{2}$ \cite{Vishwakarma:2003}.
 
Knowing the acoustic scale we can determine the location of $m$-th peak
\begin{equation}
\label{eq:47}
\ell_{m} \sim \ell_{A}(m- \phi_{m})
\end{equation}
where $\phi_{m}$ is the phase shift caused by the plasma driving effect.
Assuming that $\Omega_{\text{m},0}=0.3$, on the surface of last scattering
$z_{\text{dec}}$ it is given by
\begin{equation}
\label{eq:48}
\phi_{m} \sim 0.267 \left[ \frac{r(z_{\text{dec}})}{0.3} \right]^{0.1} =
0.267 \left[ \frac{1}{0.3} \frac{\rho_{\text{r}}(z_{\text{dec}})}
{\rho_{\text{m}}(z_{\text{dec}})} \right]^{0.1} =
0.267 \left[ \frac{1}{0.3} \frac{\Omega_{\text{r},0}(1+z_{\text{dec}})}
{0.3} \right]^{0.1}
\end{equation}
where $\Omega_{\text{b},0}h^{2}=0.02$, $r(z_{\text{dec}}) \equiv
\rho_{\text{r}}(z_{\text{dec}})/\rho_{\text{m}}(z_{\text{dec}}) =
\Omega_{\text{r},0}(1+z_{\text{dec}})/\Omega_{\text{m},0}$ is the ratio
of the radiation to matter densities at the surface of the last scattering.
 
The locations of the first two peaks are taken from the CMB temperature
angular power spectrum \cite{Spergel:2003,Page:2003}, while the location of
the third peak is from the BOOMERANG measurements \cite{deBernardis:2002}.
They values with uncertainties on the level 1$\sigma$ are the following
\[
\ell_{1} = 220.1_{-0.8}^{+0.8}, \qquad
\ell_{2} = 546_{-10}^{+10}, \qquad
\ell_{3} = 845_{-25}^{+12}.
\]
From the WMAP data, only the Hubble constant is $H_{0}=72$ km/s MPc (or the
parameter $h=0.72$), the baryonic matter density
$\Omega_{\text{b},0} = 0.024h^{-2}$, and the matter density
$\Omega_{\text{m},0} = 0.14h^{-2}$ \cite{Spergel:2003}
which give a good agreement with the observation of the position of the first 
peak.
 
In the analysis of the constraints on the bouncing cosmological model parameters
we fix the baryonic matter density $\Omega_{\text{b},0}=0.05$, the spectral
index for initial density perturbations $n=1$, and the radiation density
parameter
\cite{Vishwakarma:2003}
\begin{equation}
\label{eq:49}
\Omega_{\text{r},0} = \Omega_{\gamma,0} + \Omega_{\nu,0}
= 2.48 h^{-2} \times 10^{-5} + 1.7 h^{-2} \times 10^{-5}
= 4.18 h^{-2} \times 10^{-5}
\end{equation}
which is a sum of the photon $\Omega_{\gamma,0}$ and neutrino $\Omega_{\nu,0}$
densities.
 
Assuming $\Omega_{\text{m},0} = 0.3$ and $h=0.72$ we obtain for the standard
$\Lambda$CDM cosmological model the following positions of peaks
\[
\ell_{1} = 220, \qquad
\ell_{2} = 521, \qquad
\ell_{3} = 821
\]
with the phase shift $\phi_{m}$ given by (\ref{eq:48}).
 
From the SNIa data analysis, it was found that the Hubble constant
has a lower value. Assuming that $H_{0}=65$ km/s MPc (or $h=0.65$), we have
$\Omega_{\text{r},0} = 9.89 \times 10^{-5}$ from eq.~(\ref{eq:49}). For
further calculations we take $\Omega_{\text{r},0} = 0.0001$.
If we consider the standard $\Lambda$CDM model, with $\Omega_{\text{m},0}=0.3$,
$\Omega_{\text{b},0}=0.05$, the spectral index for the initial density
perturbations $n=1$, and $h=0.65$, where sound can propagate in baryonic matter
and photons, we obtain the following locations of first three peaks
\[
\ell_{1} = 225, \qquad
\ell_{2} = 535, \qquad
\ell_{3} = 847.
\]
We note the difference between the observational and theoretical values in
this case. We check whether the presence of the bouncing term $\Omega_{n,0}$
moves the locations of the peaks. We do not know whether it influences the 
sound velocity, but we assume that sound can propagate in it as well as in 
baryonic matter and photons.
 
To obtain the bounce, $n>4$ is necessary because the presence of the radiation
term is required by the physics of primordial plasma in the recombination epoch.
From the location of the first peak we obtain, the limit for $\Omega_{n,0}$ 
term. In the case $n=5$, with $H_0=72$ km/s MPc, we obtain that
$\Omega_{n,0}<2 \times 10^{-11}$ while for $n=6$ we have that
$\Omega_{n,0}<2 \times 10^{-17}$.
 
Please note that the special case $n=6$ was analyzed for both values 
$H_0=65$ km/s MPc and $H_0=72$ km/s MPc and the agreement with the observation 
of the location of the first peak was obtained, also for the non-zero values 
of the parameter $\Omega_{n,0}$ \cite{Krawiec:2005jj}. It means that both 
values of $H_0$ are allowed from the CMB constraints for the case $n=6$. 
However, this value is of order $10^{-10}$. The results of calculations of 
the peak locations and the values of the parameter $\Omega_{n,0}$ are 
presented in Table~\ref{tab:8}. In the special case $n=4$, the bounce term 
scale like radiation, the existing of the bounce requires 
$\Omega_{n,0} > \Omega_{r,0}$. In this case, we  also obtain the agreement 
with the observation of the location of the first peak for the non-zero 
values of the parameter $\Omega_{n,0}$ (in order to $3 \times 10^{-4}$).

Finally, we analyze the models in which we assume that sound can propagate 
only in baryonic matter and photons. With $H_0=72$ km/s MPc, in the case 
of $n=5$, we obtain that $\Omega_{n,0} < 2.2 \times 10^{-8}$ while for 
$n=6$ we have that $\Omega_{n,0} < 5 \times 10^{-14}$. For the special case 
$n=4$ we have that $\Omega_{n,0} = 2.3 \times 10^{-4}$. 

We have also calculated the age of the Universe in the $\Lambda$BCDM model. We
find that the difference in the age of the Universe is smaller than $10$ mln
years for all values of $\Omega_{n,0}$ admissible by the CMB peaks location.
So this model is admissible by the test of the age of the OHReG objects.

\begin{table}
\caption{Values of
$\Omega_{n,0}$ and location of first three peaks.}
\label{tab:8}
\begin{tabular}{c|c|cccc}
\hline
Model&Hubble constant & $\Omega_{n,0}$ & $\ell_{1}$ & $\ell_{2}$ & $\ell_{3}$ \\ \hline
Extended Bouncing model $n=4$ &$H_{0} = 65$ km/s MPc
& $ 3.0\times 10^{-4}$ & $217$ & $517$ & $816$ \\
                              &$H_{0} = 72$ km/s MPc
& $ 2.86\times 10^{-4}$ & $222$ & $526$ & $829$ \\
Extended Bouncing model $n=6$ &$H_{0} = 65$ km/s MPc
& $ 1.4 \times 10^{-10}$ & $223$ & $530$ & $847$ \\
                              &$H_{0} = 72$ km/s MPc
& $ 1.3 \times 10^{-10}$ & $224$ & $530$ & $847$ \\
\hline
\end{tabular}
\end{table}

\section{Constraint from the BBN}

The observations of abundance of light elements is in good agreement with
the prediction of the standard big-bang nucleosynthesis (BBN). It means that
the BBN does not allow for any significant divergence from the standard
expansion law, apart from the beginning of BBN to the present epoch.
Therefore, any nonstandard terms included in the Friedmann equation should
give only a negligible small change during the BBN epoch to render the
nucleosynthesis process unchanged.

It is crucial for the bouncing models to be consistent with BBN. These models
have the nonstandard term $\Omega_{n}$ which scales like $a^{-n}$ where $n>4$.
For example we analyze the cases $n=5$ and $n=6$. This additional term scales
like $(1+z)^n$. It is clear that such a term gives rise to the accelerated
Universe expansion if $\Omega_{n,0}>0$. Going backwards in time, this term
would become dominant at some redshift. If it happened before the BBN
epoch then the radiation domination would never occur and the all BBN
predictions would be lost.

The domination of the bouncing term $\Omega_{n}$ should end before the BBN 
epoch starts and we assume that the BBN results are preserved in the 
bouncing models. In this way we obtain another constraint on the value of 
$\Omega_{n,0}$. Let the model modification be negligible small during the
BBN epoch and the nucleosynthesis process be unchanged. It means that the
contribution of the bouncing term $\Omega_{n,0}$ cannot dominate over the
radiation term $\Omega_{\text{r},0} \approx 10^{-4}$ before the BBN 
($z \simeq 10^{8}$)
\[
|\Omega_{n,0}|(1+z)^{n} < \Omega_{\text{r},0}(1+z)^{4}.
\]
It means that $|\Omega_{n,0}| < 10^{-20}$ for the case $n=6$ while
$|\Omega_{n,0}| < 10^{-12}$ for the case $n=5$ respectively. Of course, the
case $n=4$ is excluded because the existence of bouncing requires in this case
$|\Omega_{n,0}| > \Omega_{\text{r},0}$, while BBN constraints require
$|\Omega_{n,0}| < \Omega_{\text{r},0}$. Let us note that inequality
$x_b \le \left( \frac{|\Omega_{n,0}|}{\Omega_{r,0}}\right)^{\frac{1}{n-4}}$
constrains the minimal size of the universe. The general conclusion from
BBN constraints is that in the present epoch, the bouncing term, if it exists,
is insignificant in comparison to the matter term.

Table~\ref{tab:9} gives the value of $z_{\text{bounce}}$ calculated for 
the best-fitted model parameters. Because the bounce should take place 
before BBN epoch so $z_{\text{bounce}} > z_{\text{BBN}} \simeq 10^8$. 
Comparing with the $z_{\text{bounce}}$ presented in Table~\ref{tab:9} 
we obtain that only two classes of models $\Lambda$CDM and $\Lambda$BCDM 
are admissible.
 
\begin{table}
\caption{The value of 
$z_{\text{bounce}}$ for the models under consideration:
Einstein-de Sitter model (CDM), $\Lambda$CDM model ($\Lambda$CDM),
bouncing model (BM), bouncing model with dust $m=3$ (BCDM)
and extended bouncing model with dust $m=3$ ($\Lambda$BCDM).}
\label{tab:9}
\begin{tabular}{c|cr}
\hline \hline
model &  $z_{\text{bounce}}$ \\
\hline
CDM             & ---   \\
$\Lambda$CDM    & ---   \\
BM              & $3.54 \times 10^4$   \\
BCDM             & $2.98$   \\
BCDM with $n=4$  & $3$   \\
BCDM with $n=6$  & $4.05 \times 10^4$   \\
$\Lambda$BCDM            & $\infty$   \\
$\Lambda$BCDM with $n=4$ & $\infty$   \\
$\Lambda$BCDM with $n=6$ & $\infty$   \\
\hline
\end{tabular}
\end{table}

\section{Conclusion}

In this paper we confront the bouncing model with astronomical observations.
We use the constraints from SNIa data, CMB analysis, and BBN and the age of
the oldest high-redshift objects.

The standard bouncing model is excluded statistically at the $4\sigma$ level.
If we take the extended bouncing model (with extra $\Omega_{\Lambda,0}$ term)
then we obtain, as the best-fit, that the parameter $\Omega_{n,0}$ is equal 
to zero which means that the SNIa data do not support the existence of the 
bouncing term in the model. We also demonstrate that BBN gives stringent
constraints on the extra term $\Omega_{n,0}$ and show that the bounce term 
is insignificant in the present epoch.

It is interesting that such bouncing models with extra inflationary expansion
are presently favored in the loop quantum approach
\cite{Bojowald:2001xe,Bojowald:2002nz,Bojowald:2003mc,Ashtekar:2004eh}.
The theory of loop quantum gravity predicts that there is no initial
singularity because of the quantum effects in the Planck scale. It is due to
the continuum break and granuality of spacetime. Therefore, we consider the
model where we assume a small positive value of $\Omega_{n,0}$ and estimate
the rest of the parameters. This model is statistically admissible. However,
when we compare this model with the standard $\Lambda$CDM model applying the
Akaike criterion, the latter is preferred.

If the energy density is so large then quantum gravity corrections are 
important at both the big-bang and big-rip. It is interesting that the 
classical theory reveals its own boundaries (i.e. classical singularities).
The account of quantum effect avoids not only an initial singularity but 
allows also to escape from a future singularity \cite{Nojiri:2004ip,
Elizalde:2004mq,Nojiri:2004pf}

The avoidance of the initial singularity arises only on the quantum ground 
because the classical theory of gravity according to the Hawking-Penrose 
theorems states that these singularities are essential if only some reasonable 
conditions on the matter content are fulfilled.
 
If we assume the classical gravity is obvious during the whole evolution of
the Universe than there is no reason to introduce the bouncing era. The 
$\Lambda$CDM model with the big-bang is a simpler model, while the bouncing 
model requires to the admittance of observationally unconfirmed assumptions.
In this way Occam's razor methodology rules out the generalized bouncing
model. The general conclusion is that the present astronomical data does not
support the bouncing cosmology.

We also adopt the methods of dynamic systems for investigating dynamics in 
the phase space. The advantages of these methods are that they offer the 
possibility of the investigation of all evolutional paths for all initial
conditions. We show that the dynamics can be reduced to a two-dimensional
Hamiltonian system. We also show structural instability of both the standard
and generalized bouncing models. Let us note that the concordance $\Lambda$CDM
models are structurally stable \cite{Smale:1980}. The structural stability is
a reasonable condition which should be satisfied by models of real physical
processes. From the dynamic investigation we obtain that all models with the
bounce are rather fragile. It means that any small perturbation of the 
right-hand sides of the dynamic equations of the model changes the topological 
structure of the phase space. The bouncing models in the space of all dynamic 
system on the plane form non-dense (zero measure) subset of this plane 
following the Peixoto theorem. Therefore, the bouncing models are untypical 
while $\Lambda$CDM models are generic from the point of view structural 
stability.

\acknowledgments 
The paper was supported by KBN grant no. 1 P03D 003 26.

\end{document}